\documentclass[conference]{IEEEtran}
\IEEEoverridecommandlockouts
\usepackage{cite}
\usepackage{amsmath,amssymb,amsfonts,nccmath,tabularx,booktabs}
\usepackage{graphicx}
\usepackage{float}
\usepackage{subfigure}
\usepackage{textcomp}
\usepackage{hyperref}
\usepackage{multicol}% http://ctan.org/pkg/multicols
\usepackage{graphicx}
\usepackage{xcolor}
\usepackage{svg}
\usepackage{algorithm}
\usepackage{algpseudocode}
\def\BibTeX{{\rm B\kern-.05em{\sc i\kern-.025em b}\kern-.08em
    T\kern-.1667em\lower.7ex\hbox{E}\kern-.125emX}}
\begin{document}

\title{fybrrStream: A WebRTC based Efficient and Scalable P2P Live Streaming Platform\\
}

\author{
\IEEEauthorblockN{Debajyoti Halder, Prashant Kumar, Saksham Bhushan, and Anand M. Baswade}
\IEEEauthorblockA{Dept. of Electrical Engineering and Computer Science \\
Indian Institute of Technology Bhilai, India \\
Email: \{debajyotih, prashantk, sakshamb, anand\}@iitbhilai.ac.in}
% \and
% \and
}

\maketitle

\begin{abstract}
The demand for streaming media and live video conferencing is at peak and expected to grow further, thereby the need for low-cost streaming services with better quality and lower latency is essential. Therefore, in this paper, we propose a novel peer-to-peer (P2P) live streaming platform, called \emph{fybrrStream}, where a logical mesh and physical tree \emph{i.e.,} hybrid topology-based approach is leveraged for low latency streaming. fybrrStream  distributes the load on participating peers in a hierarchical manner by considering their network bandwidth, network latency, and node stability. fybrrStream costs as low as the cost of just hosting a light-weight website and the performance is comparable to the existing state-of-the-art media streaming services. We evaluated and tested the proposed \emph{fybrrStream} platform with real-field experiments using 50+ users spread across India and results obtained show significant improvements in the live streaming performance over other schemes.
\end{abstract}


\section{Introduction}
Live streaming is a powerful and direct way to get connected with thousands of concurrent viewers. No matter if you are using it for your product promotion, giving a motivating speech, or just want to stream the game. It is a tool to circulate information in the form of videos. Users can enjoy the audio and video without waiting for their file to download. According to the report of Business Fortune Insights, the global video streaming market size is expected to grow from USD 342.44 billion in 2019 to USD 842.93 billion by 2027 at a compound annual growth rate of 12$\%$~\cite{fortune}. As live is on rise~\cite{liveonrise}, the demand for high-quality and lower latency services is increasing. Thus, efficient, scalable, and economical solutions are required for live video streaming.   

Today, for live streaming, media servers are being used for high quality services but at the cost of huge server deployment and maintenance expenditure. Moreover, a server is a single point of failure. Such client-server architecture is equipped to handle a lot of peers but during server failures the service might be affected if proper secondary infrastructure is not maintained. Therefore, Peer to Peer (P2P) architecture is a promising solution and has received a lot of attention. P2P technology converts the streaming consumers into the node, providing them an extra ability to forward the stream. P2P helps to remove the dependency on expensive video servers and also efficiently utilizes the upload bandwidths of the users. As the number of users increases, available resources such as bandwidth, computation power, and storage capacity also increases. So, scalability provided by P2P architecture is the one advantage with which large distributed video streaming applications can be developed in order to serve millions of concurrent users.

Web Real Time Communication (WebRTC)~\cite{webrtc} is an evolving open-source framework developed by World Wide Web Consortium (W3C) and Internet Engineering Task Force (IETF) that can be used to build real-time communication applications without installing any plugins. It provides a set of standardized JavaScript APIs for media access (getUserMedia() to access microphone and camera), media transfer (RTCPeerConnection(), and RTCDataChannel() to exchange the audio, video, and data among peers) and enables peer to peer connectivity in browsers and mobile applications. It not only eliminates the dependency of real-time communication over some third-party servers but also provides the flexibility of choosing signaling protocol (\emph{e.g.,} WebSocket, REST, etc.) to the developer~\cite{15651607}.
\setlength{\extrarowheight}{1pt}
\begin{table*}[t]
 \caption{Comparison between Tree, Mesh and Hybrid topology based approach}
\label{treevsmesh}
\begin{tabular}{ |c|p{7cm}|p{7cm}|}
\hline
\textbf{Existing methods} & \textbf{Pros} & \textbf{Cons}\\
\hline
Tree~\cite{spreadit}, \cite{6847473}, \cite{red-black}, \cite{binary-paper}
 &
\begin{itemize}
    \item Low delay
    \item Less network overhead
    \item Easy implementation
\end{itemize} & 
\begin{itemize}
    \item Low robustness and resilience during churn
    \item Need higher bandwidth for better performance
    \item Hard to maintain structure
\end{itemize} 
\\
\hline
Mesh~\cite{rkghosh}, \cite{prime}, \cite{coolstreaming} &
\begin{itemize}
    \item High robustness and resilience during churn
    \item Unstructured overlay
    \item Works well with heterogeneous bandwidth
\end{itemize} & 
\begin{itemize}
    \item High delay
    \item High network overhead (request based streaming)
    \item Complicated implementation
\end{itemize} 
\\
\hline
Hybrid~\cite{8482115}, \cite{corepeer} &
\begin{itemize}
    \item Churn resilience in mesh regions of topology
    \item Low delay and network overhead in the tree regions
    \item Works well with heterogeneous bandwidth
\end{itemize} & 
\begin{itemize}
    \item High network overhead (in the mesh regions)
    \item More complicated implementation
\end{itemize}
\\
\hline
\end{tabular}
\end{table*}

Presently, P2P live streaming systems can be broadly divided into three categories, tree-based \cite{spreadit, 6847473, red-black, binary-paper}, mesh-based \cite{rkghosh, prime, coolstreaming}, and hybrid-based \cite{8482115,corepeer}. However, the mesh and hybrid approaches suffer from the long delay, high network overhead, and complicated implementation. So they are surely not a good choice for live streaming in high load scenarios \emph{i.e.,} with a huge audience. Whereas, state-of-the-art tree-based approaches cannot handle frequent joining and leaving of peers \emph{i.e.,} churn. Therefore, we are proposing a novel hybrid approach (logical mesh and physical tree) which achieves maximum resilience during the churn with low network complicacy, latency, and overhead.\\
\noindent
\textbf{Proposed Approach:} This paper presents a hybrid approach with a logical mesh as the base network and a physical tree as the overlay network. All the user nodes are connected to each other to form the logical full mesh. This complete graph of the user nodes ensures maximum network resilience during churn or node failures which the tree-based approaches failed to cater to. The overlay network is a tree of peer-to-peer Real Time Communications (RTC) Media Channels for video and audio stream forwarding. RTC Channels are stream-push based channels for low delay and low latency streaming with low network overhead and easy implementation (this covers the issues in mesh-based approaches). Score based peer assignment enables full utilization of stream forwarding capacity of a node. This considerably decreases the height of the tree and improves end user experience. 

%Our approach can perform better than many of the well-known live-streaming platforms.
 \noindent 
\textbf{Contribution:} The contributions of the paper can be summarized as follows:
\begin{itemize}
    \item Design and evaluation of fybrrStream, a novel hybrid topology-based P2P approach for live streaming that guarantees low latency streaming and optimal peer capacity utilization with better Quality of Service (QoS).
    \item To handle the dynamics of the Internet, peer joining and peer leaving protocols  are proposed in fybrrStream.
    \item To confirm the advantages and effectiveness of fybrrStream, it is implemented and tested with real-field experiments where 50+ nodes were geographically distributed across India.
    \item The code of fybrrStream is released as research resource at \href{https://github.com/RotonEvan/fybrrstream}{https://github.com/RotonEvan/fybrrstream}.
\end{itemize}

We believe fybrrStream will bring down the server costs and will stream videos of high resolution in low latency. The rest of the paper is organised as follows. Related work is discussed in \hyperref[sec:relatedWork]{Section II}. In \hyperref[sec:systemModel]{Section III} and \hyperref[sec:protocols]{Section IV}, the proposed scheme and algorithms are discussed. In \hyperref[sec:performanceEvaluation]{Section V}, implementation of fybrrStream is presented and the results and key findings are discussed. Finally, \hyperref[sec:conclusion]{Section VI} concludes the work.
\begin{figure*}
    	\minipage{0.45\textwidth}
\includegraphics[width=\linewidth]{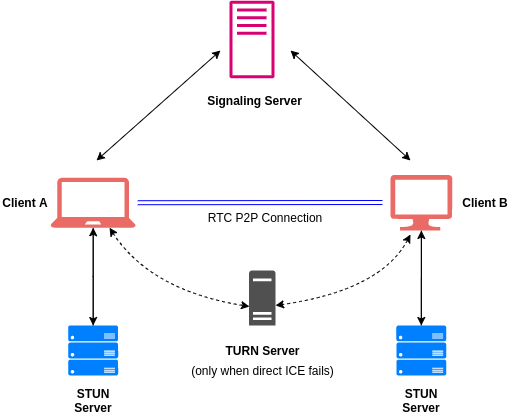}
\caption{WebRTC architecture.}
\label{fig:webrtc-arch}
	\endminipage\hfill
	~
	\minipage{0.45\textwidth}
\includegraphics[width=\linewidth]{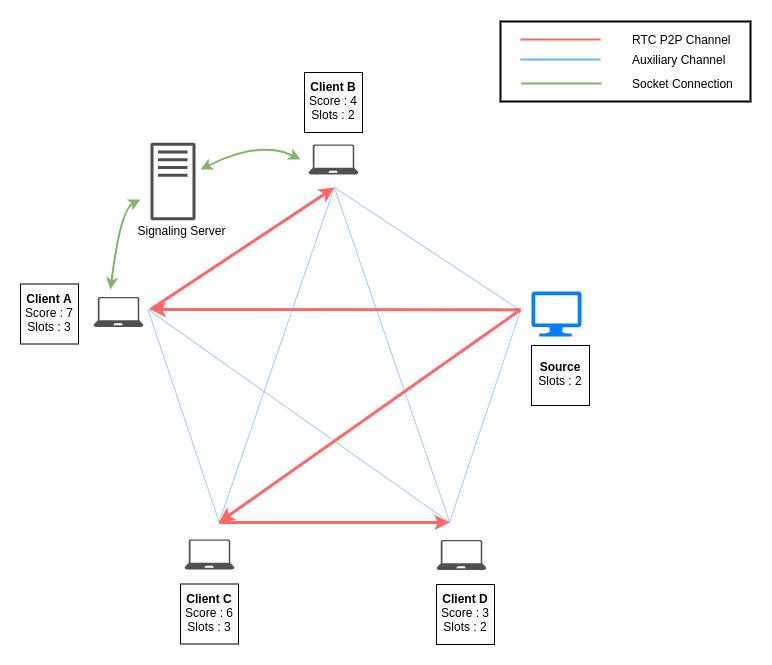}
\caption{fybrrStream architecture with logical mesh connectivity for quick recovery and physical tree-based connectivity for streaming.}
\label{fig:fybrrStream-arch}
	\endminipage\hfill
\end{figure*}
\section{Related Works}
\label{sec:relatedWork}
On the basis of P2P overlay topology, P2P streaming methods are broadly divided into tree and mesh based approaches. In a tree based approach such as in~\cite{spreadit} and Spreadit~\cite{6847473}, a node is connected to only one parent node and the stream is transmitted level by level. The disadvantage of tree-based approach is when a parent node gets disconnected, the whole subtree connected through that node gets disconnected and the data packets are lost until the subtree finds a new parent node. Also, in this level by level approach, the leaf nodes experience delay and QoS loss and the situation becomes more critical as the size of the tree increases. In the real world scenario, due to heterogeneity of the nodes, all the nodes will not have the same capacity in terms of catering streams to other peers. Some nodes may become bottleneck for better QoS and reduced delay whereas, some nodes may be underutilized. We overcome this by providing a dynamic approach in utilizing the capability of the nodes, with respect to its bandwidth, latency, and stability.

In~\cite{red-black}, the nodes are arranged in a Red-Black tree architecture so that the topology remains balanced in case of node insertion and deletion. But, under dynamic peer churn rates, the overall QoS will be highly affected because tree balancing, parent recovery and peer placement algorithms will require more time for computation.

In~\cite{8482115}, the nodes form a tree topology according to various schemes and when all the nodes have joined the session, the topology changes from tree to modified mesh as the nodes are rearranged according to their reputation level. But this approach is prone to the dynamicity of the system, as many nodes may leave the session in between voluntarily or involuntarily, some nodes may also join in between. Hence, this type of approach is not viable in case of a dynamic system where the nodes may enter or leave the session involuntarily, without informing the source/controller node.

Another WebRTC modified mesh based approach is given in~\cite{multiparty} which proposes the concept of superpeer. In this approach, a superpeer receives stream from the server and forwards it to a cluster consisting of its child nodes. The disadvantage of this approach is that the superpeer node becomes the bottleneck for better QoS and lower delay, the superpeer has to cater to many child nodes which may be slightly less competent.

In other mesh based approaches, such as,~\cite{rkghosh}, Prime~\cite{prime}, and CoolStreaming~\cite{coolstreaming}, the node asks for data packets from multiple parents and combines them to decode a stream. Also, a peer can work both as a client as well as a server, a node receives stream from its parents as a client and then forwards the stream to its children behaving as a server, this approach is called Swarming which is also used in BitTorrent~\cite{bittorrent}. 

Although tree and mesh topology are intended to minimise the server cost, they have their own pros and cons which would include low robustness and resilience during churn for tree topology and high network overhead for mesh topology. Hybrid approaches such as given in \cite{8482115}, \cite{corepeer} have also been implemented to blend the gaps created by the individual approaches, but the proposed hybrid approaches also have their own issues, such as, the high network overhead of the mesh regions and its complicated implementation. The comparison of Tree, Mesh, and Hybrid topology based approaches is given in Table \ref{treevsmesh}.

The few major issues in the previously proposed approaches are when a node is disconnected from its parent node then it also disconnects whole of its subtree from the source node, and the other, in case of mesh based approaches, a node becomes the bottleneck for receiving higher QoS for its child subtree.
Therefore, we propose fybrrStream which overcomes the network bottleneck by using a scoring mechanism that places nodes with higher capacity closer to the source node and the capacity of nodes decreases as we move away from the source node along the topology tree. This reduces the probability of a node being the bottleneck for its subtree, drastically. Also, in fybrrStream, when a node is disconnected from its parent then an auxiliary connection is triggered immediately between the neighbouring parent nodes. This is discussed in more detail in the following section.
\section{Proposed Architecture}
\label{sec:systemModel}
\setlength{\extrarowheight}{1pt}
\begin{table*}
 \centering
 \caption{Definitions of Terms used}
 \label{table:term-def}
\begin{tabular}{ |p{2cm}|p{15cm}|}
\hline
\textbf{Term} & \textbf{Definition} \\
\hline
\hline
RTP Connection &
Real-time Transport Protocol Connection between two users for streaming video and audio over IP networks.
\\
\hline
Peer/Node/Client
 &
A user who is connected to another user through an RTP Connection.
 \\

\hline
Room &
A private live streaming web-link on fybrrStream web application.
\\
\hline
Score &
A numerical value calculated when a node joins a fybrrStream room which signifies stream forwarding capability of that node.  %stream forwarding capability of 
\\
\hline
Slots/Capability &
Number of more peers a node can accommodate as a child, \emph{i.e.,} number of peers to forward the stream to.
\\
\hline

Auxiliary Connection &
Temporary RTP Connection between an orphan node and a temporary parent until a new more \emph{capable} parent is assigned.
\\
\hline
Number of Hops &
The number of times a stream originating from the source node is forwarded through intermediate nodes to reach the destination~node.
\\
\hline
\end{tabular}
\end{table*}
% For live streaming, media servers are being used for high quality services but at the cost of huge server functional and maintenance expenditure. Moreover a server is a single point of failure. This client-server architecture is equipped to handle a lot of peers but during server failures the service might be affected if proper secondary infrastructure is not maintained. Thus, P2P architecture has received a lot of attention.
% fybrrStream is developed for a scenario where the source of the live stream needs to broadcast to a multitude of viewers. Live broadcast streaming can be seen in video conferences (On platforms such as Google Meet, Cisco Webex, Zoom) where one user might broadcast a webcam video or a presentation to multiple users. Over-the-Top (OTT) services, live programmes broadcast are also examples of the same scenario. While OTT streaming is managed by Content Delivery Networks (CDNs) and can be tweaked in quality as per user network constraints, live streaming programmes need to be delivered with the lowest latency possible and with minimum packet loss.
% \subsection{System Architecture}
% \subsection{Hybrid Overlay Network - Best of both Worlds}

In this section, we have described the fybrrStream system architecture and all the different components of the architecture. 

% \begin{figure}[H]
%     \centering
%     \includegraphics[scale=0.33]{intro.png}
%     \caption{WebRTC Architecture}
%     \label{fig:webrtc-arch}
% \end{figure}
%\section{Proposed Scheme}
%\label{sec:proposedModel}
\textbf{System architecture: }fybrrStream uses WebRTC architecture for peer-to-peer streaming. The WebRTC architecture (see Fig.~\ref{fig:webrtc-arch}) consists of the following components:
\begin{enumerate}
    \item \textbf{Signalling server} sets up Real-time Transport Protocol between two nodes (a parent-child relation).
    \item \textbf{Session Traversal Utilities for NAT (STUN) servers} retrieve public IP of a node.
    \item \textbf{Traversal Using Relay around NAT (TURN) servers} are backups for RTP Connections.
\end{enumerate}

fybrrStream has in background a logical full mesh network and a physical tree network of nodes for communication. The full mesh network is a network of user nodes connected to each other. In case of a node failure, auxiliary connections are created between the affected nodes to ensure uninterrupted streaming. The physical overlay tree network is a tree of nodes connected with RTC Media Channel links. These links are used for peer-to-peer stream forwarding. A score based peer assignment algorithm is followed to ensure full utilization of stream forwarding capability of a node. The peer joining protocol at the signalling server ensures quick joining of nodes, peer leaving and restructuring protocol ensures swift fault detection and network restructuring. Fig. \ref{fig:fybrrStream-arch} shows the fybrrStream architecture. All the components of the architecture have been explained below in detail. Table \ref{table:term-def} defines all the terms that are used multiple times further in this paper.
% \begin{figure}[H]
%     \centering
%     \includegraphics[scale=0.3]{score.png}
%     \caption{fybrrStream Architecture}
%     \label{fig:fybrrStream-arch}
% \end{figure}

\subsection{Signalling Server}
A signalling server is used to establish RTC Connection between peers. Connection establishment includes exchange of SDP (Session Description Protocol) and ICE (Interactive Connectivity Establishment) candidate. If the ICE candidate connects successfully then RTC Connection is setup for stream delivery. The SDP contains the session description of the two peers, with configuration of the media to be sent by the source peer and the response description of the receiver peer. 
\subsection{STUN/TURN servers}

To setup RTC Connection, STUN servers are used to retrieve public IP of a device for initialising RTP over User Datagram Protocol (UDP) . WebRTC uses UDP for stream delivery. If the ICE candidate fails to connect due to any possible reason (if a device is behind symmetric Network Address Translation (NAT), or is protected by a firewall which discards UDP data packets), then Traversal Using Relays Around NAT (TURN) servers are used to relay the stream around NAT.

\subsection{Overlay Network of RTP Connections}
When a new RTC Connection is established between an old and a new peer, the new link extends the overlay network which actually is a tree. The predefined rules of peer selection (request-based protocol for parent assignment) protocol says that every node in a level must have at-least 2 children until the level is complete. This makes the worst case complexity for a packet to reach any node from the source $\mathcal{O}(\log{}n)$ for a total of $n$ nodes in the network. 

\subsection{Fault-proof Architecture}
Our proposed architecture is using WebRTC to make connections and transfer stream between the peers. Internet Connectivity Establishment (ICE) service will find the possible candidates to connect to a peer. Although we are proposing a distributed model for live streaming but maintenance of overlay topology is taken care of by a centralised entity \emph{i.e.,} signalling server. New peers will be added to the topology according to their score and available slots as we know that nodes can be heterogeneous in terms of internet connectivity and thus have varying capacity.

Video stream will be pushed from the parent node to their children in this tree overlay topology. In case, children fail to receive the stream (\emph{e.g.,} parent node failure), they utilise auxiliary connections to continue the stream.

\subsection{Auxiliary Connections}
\label{Sub:Aux}
The base full mesh topology is utilised to the fullest to provide seamless stream delivery. When a peer-to-peer RTC Connection in the overlay network is broken, then an auxiliary peer starts sending stream to the orphaned nodes temporarily until the normal overlay is established. The server finds that a node has failed. It then informs every orphaned child so that they can create an auxiliary RTC Connection with an auxiliary peer to minimise the interruption in service.

The auxiliary links are established with the peer that can provide the stream with the least number of hops possible. WebRTC enables the quickest transition between stream sources. The auxiliary peers will be stored for every node such that there is no lag in the transition. Siblings of parent and grandparent will be the preferred for auxiliary nodes. The list of auxiliary nodes will be sorted to get the best auxiliary node. The list is required to be kept prepared with the best node identified.

\subsection{Summary of proposed architecture}

% We have developed a hybrid approach where our base topology is a full mesh (a complete graph of all the peers as nodes) and the overlay network is a tree of Real-Time Communications (RTC) Connections. Our approach has mobility of an unstructured mesh during a churn and also the low latency stream delivery of a push-based stream in a tree topology. The overlay network is created by a request-based protocol like in a mesh topology. This gives a newcomer peer the best parent to receive stream from. During network churns, the links in the full mesh become auxiliary RTC Connections to deliver the stream in case of multiple node failures. WebRTC implementation has optimal bandwidth requirements and has been tested to work under low bandwidth network conditions as well.

The hybrid approach constitutes of a full mesh based topology and an overlay tree network of RTC Connections. The reason for going with this approach is predominantly to lower the reconnection delay after a network abruption. RTC Media Channels are very fast to establish in WebRTC. During network churns, the links in the full mesh become auxiliary RTC Connections to deliver the stream in case of multiple node failures. WebRTC implementation has optimal bandwidth requirements and has been tested to work under low bandwidth network conditions as well. Our approach has mobility of an unstructured mesh during a churn and also the low latency stream delivery of a push-based stream in a tree topology. The overlay network is created with the best parent-child RTC connections with a global outlook on the current network conditions. No peer is over-populated with children nodes if there already exists another peer with unused~capability.

\section{fybrrStream Protocols}
\label{sec:protocols}

In this section, we describe the protocols followed during peer assignment, peer joining and peer leaving. Also, special scenarios for peer joining and peer leaving have been given as examples to describe the protocols followed during those circumstances. 

\subsection{Score based Peer Assignment}
Once a source peer has joined the network and it is willing to stream the media, parents will be assigned to new peers on their respective joining as per the suggested Joining Protocol (See Section \ref{Sub:join} for details). The score of a node is calculated after taking uploading bandwidth, video streaming rate, latency, and stability of that node into consideration. Eqn.~(\ref{eq:score}) has been proposed for the calculation of score.

\begin{multline}\label{eq:score}
Score(peer) \; =\; k_{1}*\Bigg(\frac{upload\_bandwidth}{streaming\_rate} \Bigg)\\
+ k_{2}*\Bigg(\frac{1}{latency}\Bigg) + k_{3}*\Bigg(\frac{active\_duration}{num\_of\_failure}\Bigg)
\end{multline}

First term in Eqn.~(\ref{eq:score}) represents the serving capacity of a peer or the number of children a particular peer can forward stream to without much impact on its own performance. $upload\_bandwidth$ is the upload bandwidth of the node and $streaming\_rate$ is the rate at which the data packets are being generated, both given in Mbps. Next term depends on the time taken for the stream to reach from source to the peer. $latency$ is the latency observed on the node when establishing the connection and for further data packet transmission. Last term signifies the stability of the peer \emph{i.e.,} the measure of node failures. $active\_duration$ is the time duration for which the node has been active and $num\_of\_failure$ is the number of times the node has failed or lost connection. The second and third term is used to update the score of a peer during the stream. Where, $k_{1}$, $k_{2}$, and $k_{3}$ signifies the weights of the terms and values of these constants varies according to applications. The weights sum up to one, \emph{i.e.,} $k_{1} + k_{2} + k_{3} = 1$.

Also, the peer is assigned slots \emph{i.e.,} the number of children nodes it can handle. The slots are assigned according to the upload bandwidth of the peer. Slots of a peer $P$ can be calculated using Eqn.~(\ref{eq:slots}).

\begin{equation}\label{eq:slots}
slots_{P} \; = \; \frac{upload\_bandwidth_{P}}{streaming\_rate}
\end{equation}

\begin{algorithm}[H]
\caption{Peer Joining Algorithm}
\label{Algo:Join}
\begin{algorithmic}[1]

\Procedure{Peer Joining - Server Side}{$room, score, slots, peer$}
    \If {room exists}
        \State Add new node into room
        \If {slots(source) $>$ 0}
            \State Parent[peer] $\leftarrow$ source
        \Else
            \State minSlotPeer = findMinSlotChildOfSource() \Comment{Find child with lowest number of available slots from the children of source}
            \If {slots(minSlotPeer) $>$ slots(peer)}
                \State Parent[peer] $\leftarrow$ source
                \State Parent[minLimitPeer] $\leftarrow $ peer
            \Else
                \State bestParent $\leftarrow$ findBestParent() \Comment{Find peer with highest score and non-zero available slots}
                \State Parent[peer] $\leftarrow$ bestParent
            \EndIf
        \EndIf
    \Else
        \State Create a new room
        \State source $\leftarrow$ peer
    \EndIf
\EndProcedure

\end{algorithmic}
\end{algorithm}

\subsection{Peer joining protocol}
\label{Sub:join}
Whenever a new peer wants to join the streaming, score and available slots will be calculated on the client side with the help of the Eqn.~(\ref{eq:score}). It will then send a “JOIN” request to the signalling server. Score and number of available slots will be embedded in this joining request only. Server finds the most suitable peer to be the parent of the new peer based on the score and available slots of the peers that are already present in the network. Newly joined peer can even replace a direct child of source if it has a better score than that node in terms of forwarding capabilities and serving power. Algorithm \ref{Algo:Join} is the pseudo-code for the Peer Joining Algorithm. The algorithm is explained further with few possible joining scenarios as examples.
$\;$\\ \textbf{Scenario 1 :} New peer joins. Room contains none except Source.

 New Peer A has joined the network and has initiated Stream receiving protocol by sending ``JOIN" request to signalling server. Server recognises Peer A being the only node in the room except Source and connects Peer A with Source directly. An RTC Peer Connection Channel is setup between Source and Peer A.
\begin{figure}[H]
    \centering
    \includegraphics[scale=0.4]{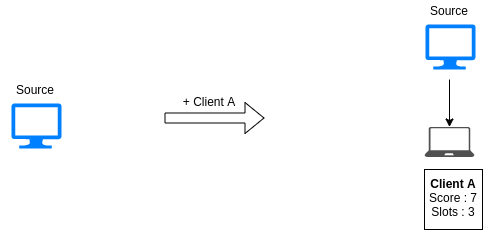}
    \caption{New client node addition to network after Source}
    \label{fig:my_label}
\end{figure}$\;$\\
\textbf{Scenario 2 :} New peer joins. Source has children with lesser capacity.

Source with two children A and B with capacity 3 and 4, respectively, are present in the room along with respective sub-trees. New Peer C joins the network. C has a capacity of 6. Server finds the child of Source with the minimum capacity lesser than that of Peer C using the function \textit{findMinSlotChildOfSource()}. In this case server finds Peer A. It makes Peer C the direct child of Source and Peer A child of Peer C. This is done to keep high capacity nodes in the front line of the overlay such that stream delivery has the least latency possible.
\begin{figure}[H]
    \centering
    \includegraphics[scale=0.25]{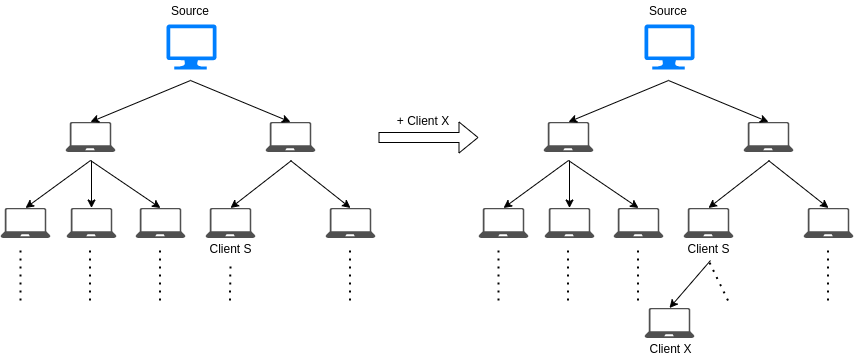}
    \caption{New node replaces old child of source}
    \label{fig:my_label}
\end{figure}

\textbf{Scenario 3 (General scenario):} New peer joins. Source has children, all with higher capacity than new peer.

Joining request is sent to the signalling server from a peer X with score and slots values. Server checks that the source peer is working at its full capacity and none of its children capacity is less than the new peer. As per our proposed algorithm, signalling server finds the peer with highest score S using the function \textit{findBestParent()} which is available to server new peers \emph{i.e.,} currently not working at its full capacity and assigns it to new node. Peer X can start receiving streams from the assigned parent.
\begin{figure}[H]
    \centering
    \includegraphics[scale=0.3]{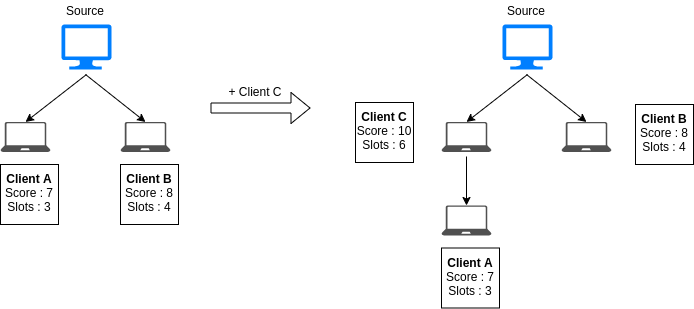}
    \caption{New client node addition to network}
    \label{fig:my_label}
\end{figure}
\subsection{Peer leaving protocol}
A peer has left the streaming network can be concluded from two ways. Firstly, if the server has received a ``LEAVE” request from the peer. Secondly, a heartbeat ping-pong protocol runs on the server at regular intervals. If a server misses a pong message of a peer that would signify a node failure. In both the cases, leaving peer will be replaced by its best child and will then be removed from the network. If leaving peer was a free rider and its parent was the source node, then replace it with the most suitable peer (any node from the grandchildren sub-trees of Source) in the overlay network for efficient utilisation of uploading bandwidth of source and to maintain maximum width tree topology. 
\begin{algorithm}[H]
\caption{Peer Leaving Algorithm}
\label{Algo:Leave}
\begin{algorithmic}[1]

\Procedure{Peer Leaving-Server Side}{$peer$}       
    \State parent $\leftarrow$ Parent[peer]
	\If {length(children[peer]) is 0} \Comment{Leaving peer was a free rider}
	    \If {parent is source} 
        	\State bestPeer $\leftarrow$ findNextBestNode() \Comment{Find a peer with maximum score which is not a current child of source.}
        	\State Parent[bestPeer] $\leftarrow$ source
        \EndIf
    \Else
        \State bestChild $\leftarrow$ getBestChild(peer) \Comment{Find most suitable child of leaving node to replace it}
        \State Parent[bestChild] $\leftarrow$ parent
    \EndIf
\EndProcedure
\end{algorithmic}
\end{algorithm}

Till the new parent is assigned to the orphaned children (if any), these children will continue to receive the stream from the auxiliary connections (as explained earlier in Section \ref{Sub:Aux}) to avoid any glitches that might occur because of peer leaving. Algorithm \ref{Algo:Leave} is the pseudo-code for the Peer Leaving Algorithm and has been explained with possible scenarios as examples.
$\;$\\
    \textbf{Scenario 1 :}  Node  with non-zero children fails and leaves. Node A was an intermediate node in the overlay network.
    
    Server will find the child with the maximum score from the set of direct children of the leaving node. Maximum score child will now receive the stream from the parent of leaving node and the rest of the children of leaving node will join the network as per the joining protocol.\\
    \noindent
    \textbf{Scenario 2 :} A leaf node fails or leaves the network. Node was also a direct child of Source.
    
    Leaving node will create a vacancy of a direct child of source. Signalling server will find a peer with the maximum score which is not a current child of source and will put it (keeping subtree of this node  intact) directly under the source to fill that vacancy.
\begin{figure}[H]
    \centering
    \includegraphics[scale=0.28]{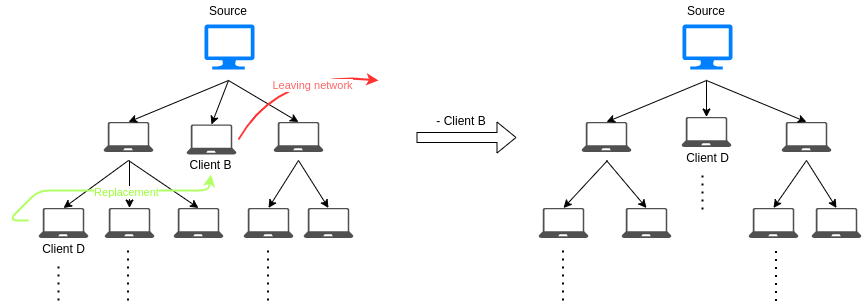}
    \caption{A direct child of Source leaves the network.}
    \label{fig:leave}
\end{figure}
\subsection{Peer Failure}
Signalling server will detect peer failure with the unacknowledged heartbeat pings. Once any failure is detected, node leaving protocol will be executed with only one difference of incriminating node failure count by one every time it fails. Node failure count will have an impact over the stability value of node and so score also changes in the similar fashion. New score will be recalculated and used only when this node again joins the streaming network.

\section{Performance Evaluation}
\label{sec:performanceEvaluation}
\begin{figure}[t]
    \centering
    \includegraphics[scale=0.18]{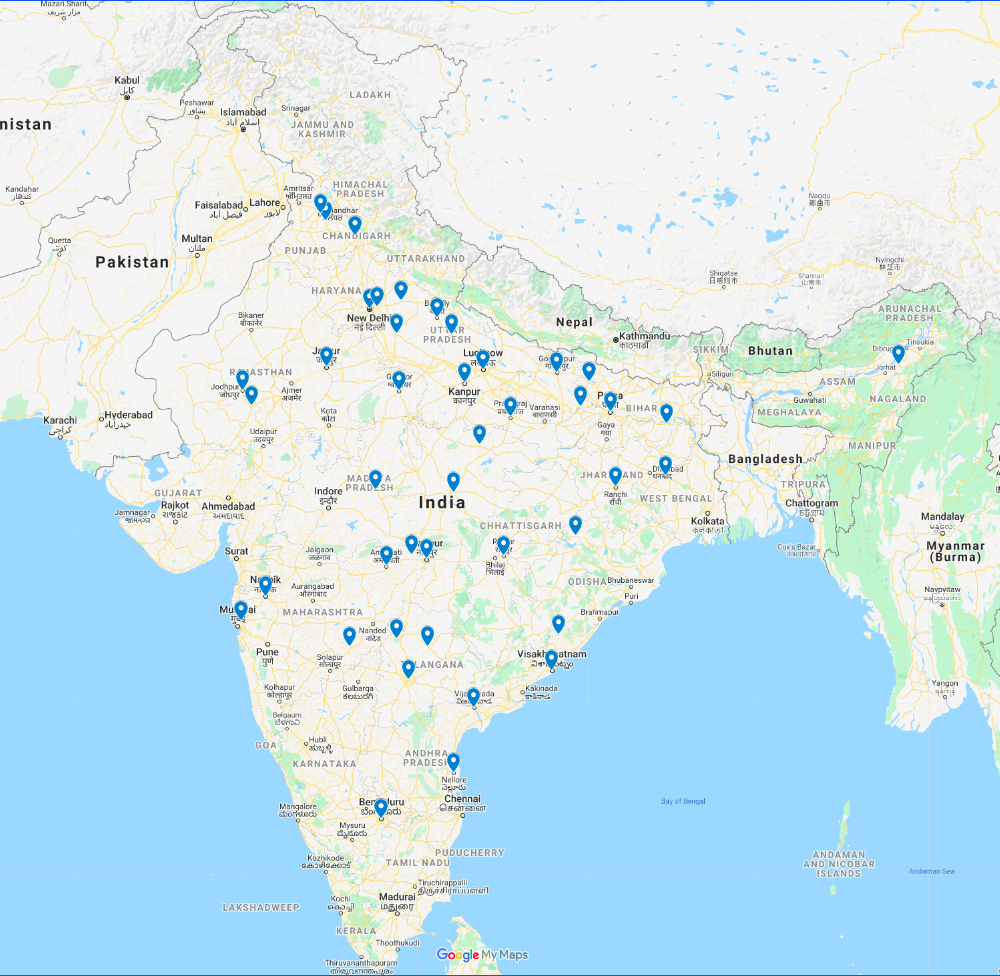}
    \caption{Geographical distribution of peers during fybrrStream field test.}
    \label{fig:map}
\end{figure}

% write something here
%In this section, we evaluate our proposed model against some performance metrics to showcase the benefits and performance gains achieved through our system. 
We tested fybrrStream in a country-wide live streaming session. Total 56 students of Indian Institute of Technology Bhilai, India joined the field test. These students were at their homes and their geographical locations are spread across India as shown in Fig. \ref{fig:map}. The farthest geographic distance of a peer from source node was around 1700 km. We live streamed a YouTube video \cite{youtube} in high-definition 1080p resolution. 
%\textbf{More about the experiment setup is written in Section \ref{Sub:results}.}
%We compare the fybrrStream scheme with a binary tree approach which is given in \cite{red-black}. In comparison to this fybrrStream has smaller tree height for the same number of nodes in the session as shown in Section \ref{Sub:results}. We compare fybrrStream also with a quaternary tree approach which has smaller tree height than fybrrStream, but looses on other parameters such as latency and jitter. Thus, the performance evaluation suggests that fybrrStream is the optimal approach.
\begin{figure*}[t]
    	\minipage{0.32\textwidth}
        \includegraphics[width=\linewidth]{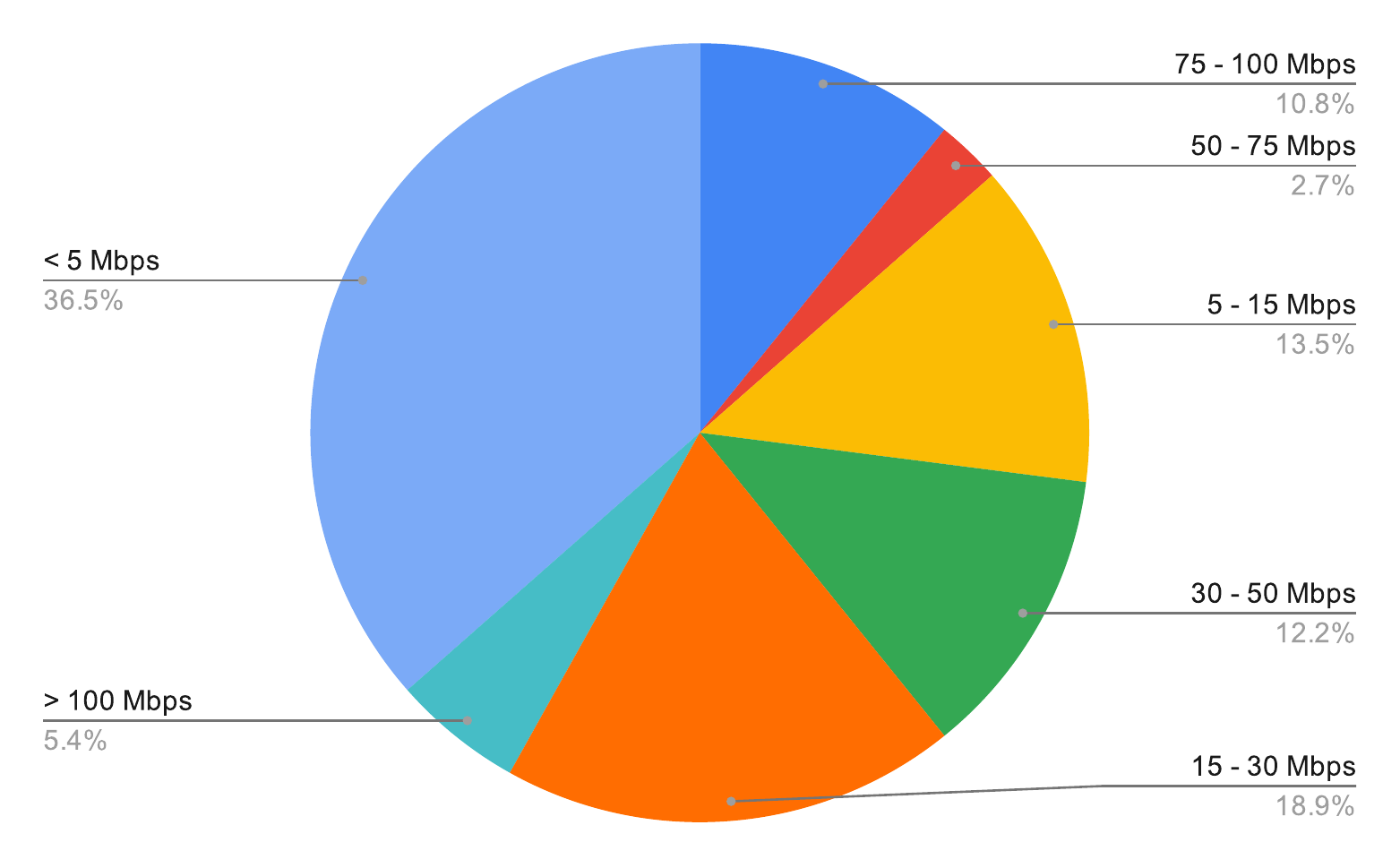}
        \vspace{0.2cm}
        \caption{Upload network bandwidth distribution for users obtained using Ookla or Google~Speedtest.}
        \label{fig:band_dist}
    	\endminipage\hfill
	~
		\minipage{0.32\textwidth}
        \includegraphics[width=\linewidth]{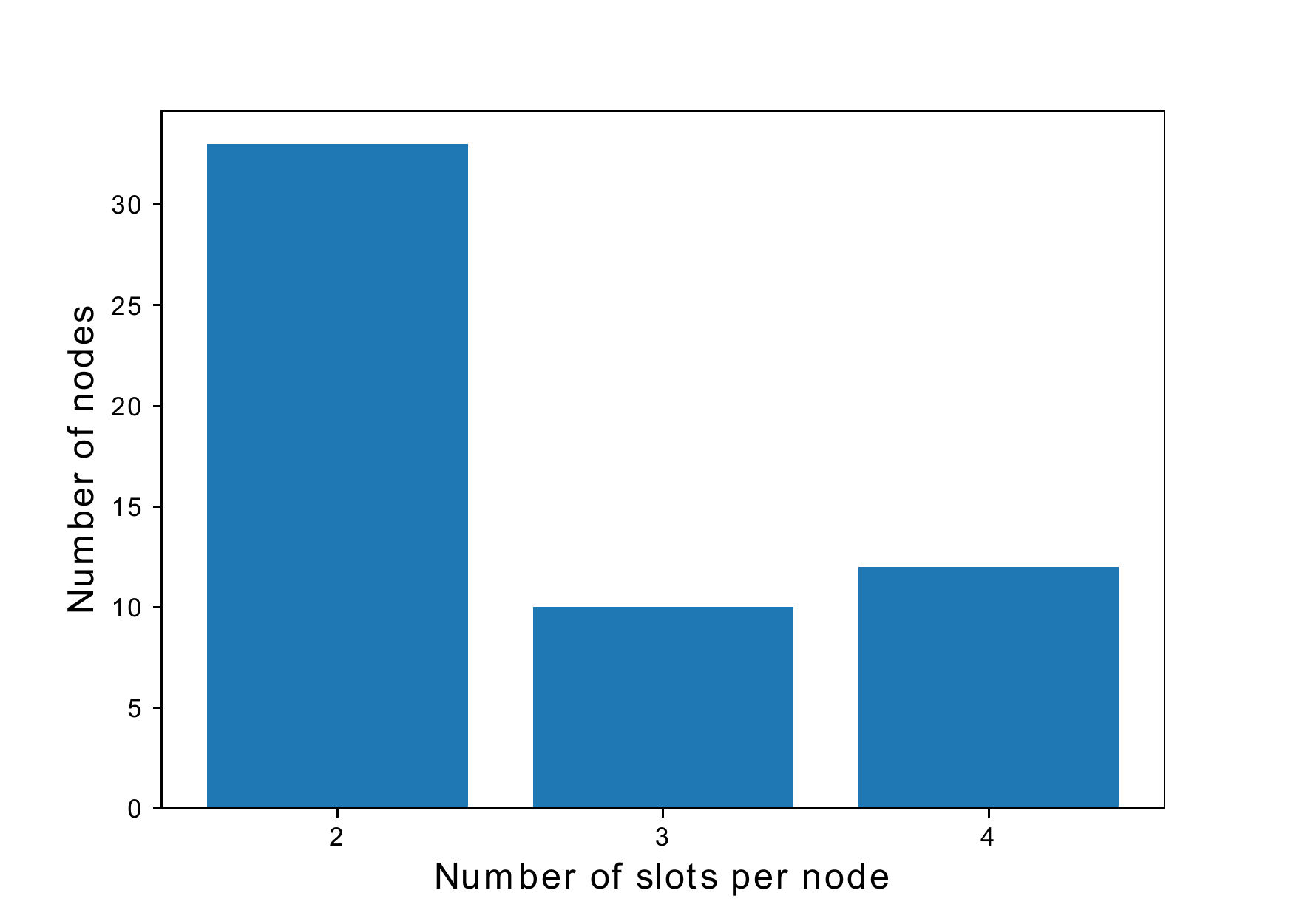}
    \caption{Plot for number of nodes with respect to the number of slots.}
    \label{fig:slot_dist}
    	\endminipage\hfill
    		~
		\minipage{0.32\textwidth}
          \includegraphics[width=\linewidth]{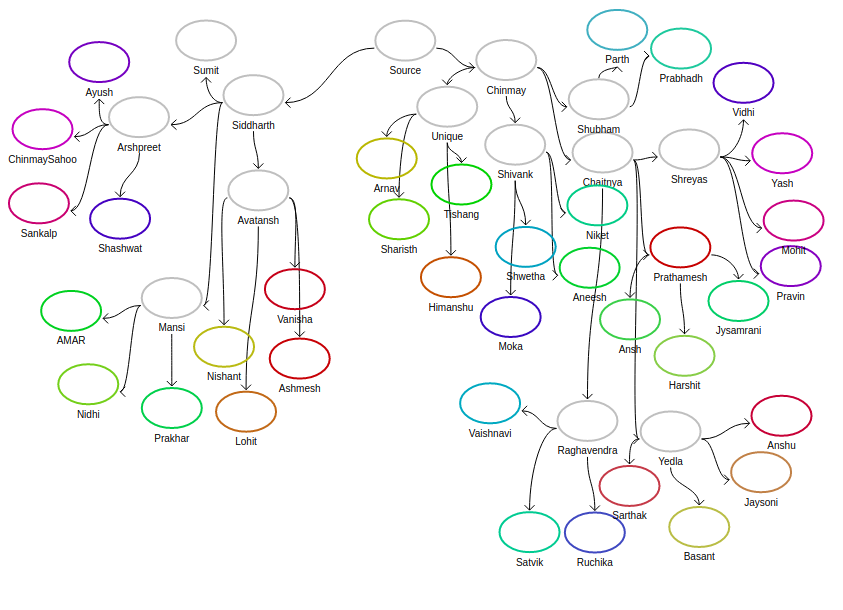}
    \caption{An instance of topology in case of fybrrStream during the live field test.}
    \label{fig:fybrr_topo}
    	\endminipage\hfill
\end{figure*}
\subsection{Performance Metrics}
We designed fybrrStream with main focus on reduced latency, packet loss, and startup delay. Therefore, to evaluate our approach and to assess the impact of our approach on the mentioned parameters, we define performance evaluation metrics as
\begin{itemize}
    \item[-] \textbf{Packet delivery ratio (PDR):} It is the ratio of the number of packets received by a node to the maximum number of packets that could have been received.
    \item[-] \textbf{Latency:} It is the time difference in receiving the packet by a node after it was uploaded by the source node. Calculated as defined in Section 6.4.1. of RFC 3550 \cite{rfc3550}.
    \item[-] \textbf{Jitter:} It is the variation in inter-arrival time between consecutive data packets. Calculated as defined in Section 6.4.1. of RFC 3550 \cite{rfc3550}.
    % \item[-] \textbf{Capability utilization:} It is the ratio of the slots taken up by the nodes to the total number of slots available in the topology, averaged over duration of the whole session.
    % \item[-] \textbf{Node Instability:} It is the ratio of number of involuntary disconnections from the network to the total duration of active state of the node.
    \item[-] \textbf{Startup delay:} It is the time taken by a node, after joining the session, to start receiving the stream.
    % \item[-] \textbf{Auxiliary Connection Utilization:} It is the ratio of number of Auxiliary connections used by the nodes to the number of original parent of all the nodes.
    %
\end{itemize}
\subsection{Experimental Setup}
\label{Sub:setup}
Implementation of fybrrStream is hosted on Heroku~\cite{heroku} servers and the same was used for the experiments performed. Performance of proposed approach and algorithms are analysed in a real-life scenario with a field-test. In this field test, around 56 students from our institute participated in the performance evaluation as an individual peer from their home, thus we were able to get such diverse geo-location (as shown in Fig.~\ref{fig:map}) of the peers. The nodes were heterogeneous in nature with respect to Internet speed and computation power. This was helpful in showcasing a more realistic scenario where the nodes are on different networks, experience different Internet bandwidths and are using different mediums for communication (Ethernet, WiFi, 3G, 4G) with different signal strengths. Both mobile and desktop users were allowed to participate. Institute TURN servers were utilised to relay stream in case direct peer connection was not possible because of firewall or symmetric NAT. Our participants were distributed across the map of India as shown in Fig. \ref{fig:map}, out of which few peers shared the same city/town/village.
The source node is responsible for creating a room and sharing the stream to other participants in the room. The users joined the room through a web-link which was provided to them prior to the experiment. The source node started the stream after all the users joined the room to maintain consistency in the collected data. As the users were joining the room, fybrrStream was simultaneously constructing the peer topology and measuring the startup delay for each joining node. WebRTC statistics was accessed with the help of getStats() API~\cite{getStats} which returns a WebIDL dictionary~\cite{webidl} containing the values of several monitored metrics at a specific moment in time. During the live stream, we collected data for packet loss, packet received at each node, along with it, jitter and latency is calculated for each leaf node of the constructed overlay tree topology. 
%\begin{figure}[H]
 %   \centering
  %  \includegraphics[scale=0.4]{num_slots.eps}
   % \caption{Plot for Number of nodes with respect to the number of slots.}
    %\label{fig:slot_dist}
%\end{figure}

Fig.~\ref{fig:band_dist} shows the upload network bandwidth distribution of participated users in the field test. The values are in megabits per second obtained through Ookla Speedtest or Google Speedtest. Fig.~\ref{fig:slot_dist} shows the number of slots assigned to the number of nodes present in the session and from the plot it can be seen that the maximum number of nodes have 2 slots whereas some nodes have 3 and 4 slots. For this experiment, we have restricted the number of slots to 4 due to network and computational limits on the user-end.
%  One peer was streaming a video for around 30 minutes and participants were asked to access/join the streaming from the hosted site. For a better evaluation, performance metrices were computed, stored and then downloaded in the client side only. Generated log files were then processed further to get some remarkable insights from the experiment. Fybrrstream is compared with the other two tree based schemes. In first, a peer can forward the stream to maximum of 2 children i.e. overlay network is a binary tree termed as 'binary' for the further discussion of the paper. Quaternary tree is the overlay topology in second scheme.
%\begin{figure}[H]
 %   \centering
  %  \includegraphics[scale=0.25]{Topology_1.png}
   % \caption{An instance of topology in case of fybrrStream during the live field test.}
    %\label{fig:fybrr_topo}
%\end{figure}
Fig. \ref{fig:fybrr_topo} shows the overlay network topology formed by the nodes in the fybrrStream scheme. As can be clearly seen in the figure that the source node is connected to 2 nodes, which means the source node has 2 slots which is then followed by the 2 children nodes of the source node, with 4 slots each. It can also be observed that the nodes with higher value of slots are placed nearer to the source node compared to the nodes with lower number of slots. This figure represents one instance of the whole session, as the topology varies slightly with each node joining and leaving mid-session.

\subsection{Experimental Results}
\label{Sub:results}
Table~\ref{table:params} defines the parameters and values considered for the experiment.
\setlength{\extrarowheight}{1pt}
\begin{table}
 \centering
 \caption{Experimental Parameters}
 \label{table:params}
\begin{tabular}{|p{2.1cm}|p{5.4cm}|}
\hline
Parameter             & Values                                                                      \\ \hline \hline
Bandwidth range       & 5 - 100 Mbps                                                               \\ \hline
Slots range           & [2, 3, 4] slots for [5-15, 15-50, 50-100] Mbps                                                                       \\ \hline
Number of nodes       & 57 (including source node)                                                  \\ \hline
Stream quality        & 1080p                                                                       \\ \hline
Frames per second     & 20 fps                                                                      \\ \hline
Browser support       & Chrome, Firefox                                                             \\ \hline

Compression algorithm & VP8, VP9 (Chrome)\newline H.264 (Firefox) \\ \hline
Stream video          & "The story of an idea" - Google IO 2017 | Google Developers                 \\ \hline
Video platform        & YouTube                                                                     \\ \hline
\end{tabular}
\end{table}
The proposed scheme is compared with two other tree-based schemes - 'binary tree' (2 children nodes per peer) and 'quaternary (quad) tree' (4 children nodes per peer). The reason behind choosing these schemes to compare with fybrrStream  was the similarities between the schemes (binary and quad are tree based, and fybrrStream has a tree overlay network) but the heuristics causing massive differences in the results. Binary tree has been suggested in multiple papers~\cite{red-black}, \cite{binary-paper}. We also wanted to reduce the height of the tree to decrease latency in the leaf nodes. So we considered a quad tree, in which every peer would have 4 children nodes (regardless of bandwidth constraints), and that would decrease the height of the tree.
\begin{figure}[H]
    \minipage{0.23\textwidth}
    \includegraphics[width=4.6cm]{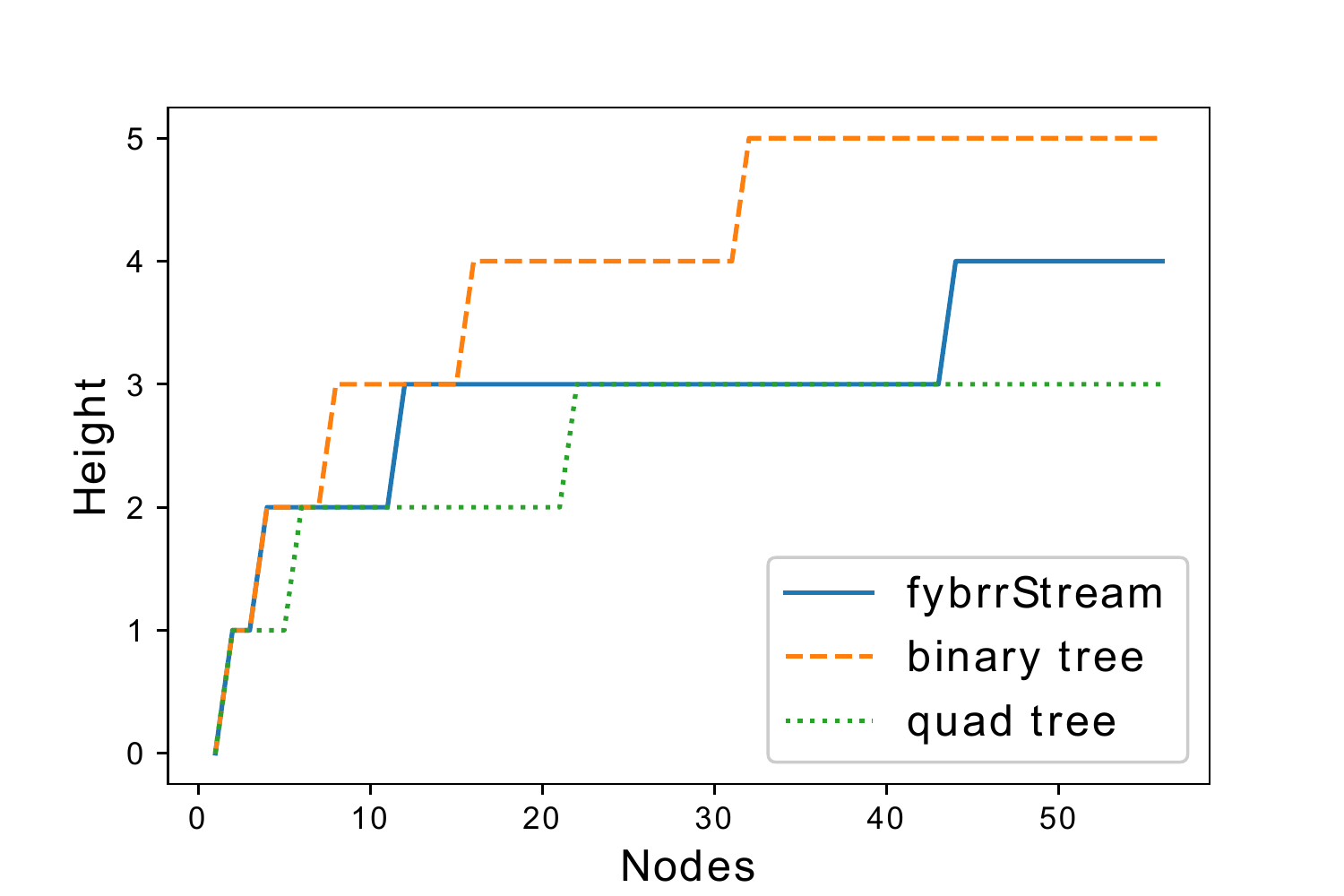}
    \caption{Height of topology tree as the nodes enter in the room.}
    \label{fig:hop}
    \endminipage\hfill
	~
	    \minipage{0.23\textwidth}
       \includegraphics[width=4.6cm]{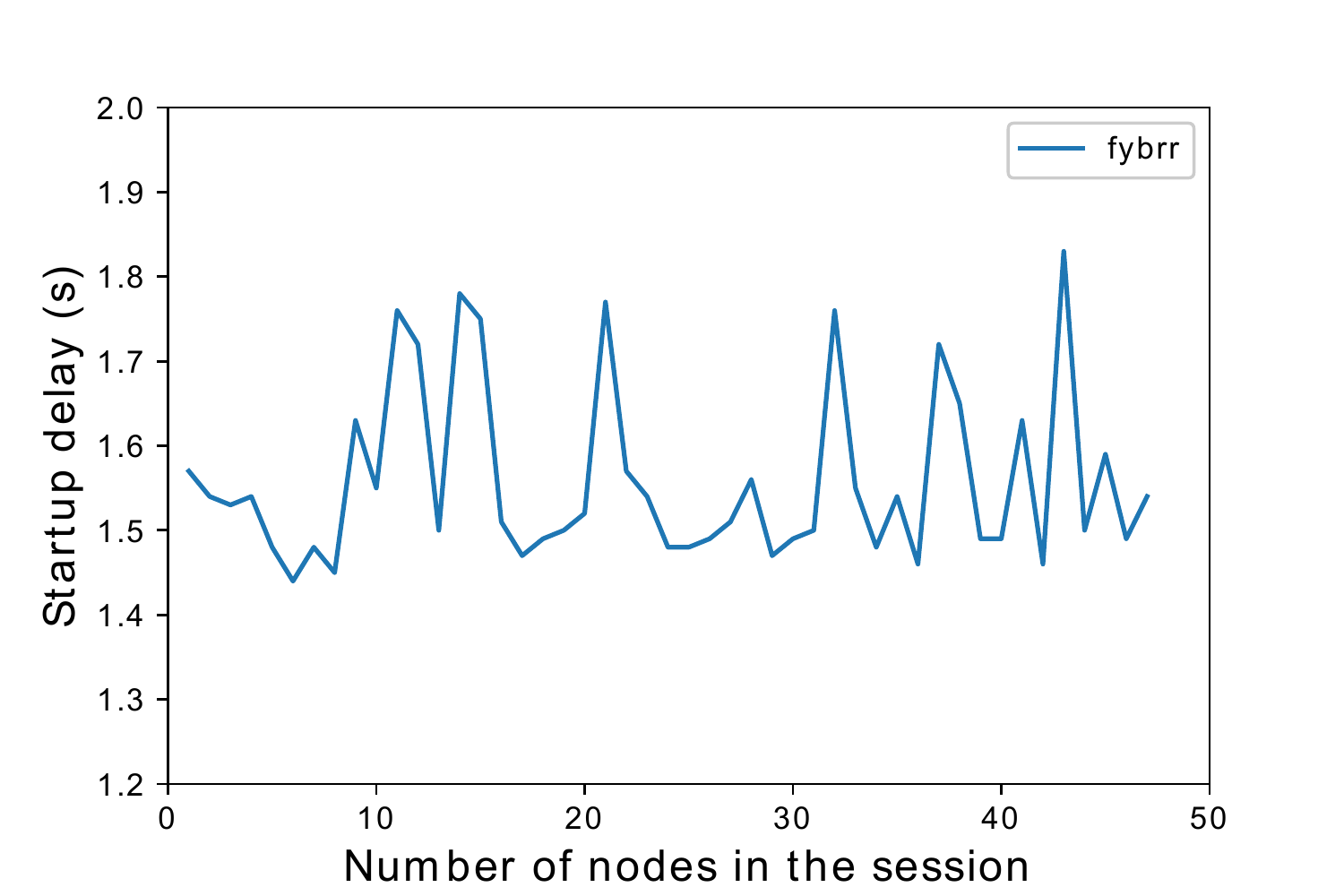}
    \caption{Startup delay (s) vs number of nodes in the session.}
    \label{fig:startup-delay}
    \endminipage\hfill
\end{figure}
 The height of the topology tree with increasing number of nodes for different schemes is shown in Fig.~\ref{fig:hop} and can be related through Eqn.~(\ref{eq:hops}):

\begin{equation}\label{eq:hops}
height_{quad}\leq height_{fybrrStream}\leq height_{binary}
\end{equation}
where $height_{quad}$, $height_{fybrrStream}$ and $height_{binary}$ is the height of the topology tree in case of quad tree, fybrrStream and binary tree, respectively.
In fybrrStream, peers are assigned the maximum number of children nodes they can support based on their upload bandwidth. For this experiment, we considered to keep a minimum of two slots for every peer as all had an upload bandwidth of around 5 Mbps, so a peer could afford forwarding the stream to atleast 2 peers. Therefore in the worst case an overlay network created will have been the same as that created in the binary tree scheme.

Now, if $n_{1}$ nodes have $2$ children nodes assigned to them, $n_{2}$ nodes have $3$, and so on till $n_{i}$ nodes have $i+1$ children nodes assigned to them, and 

\begin{equation}
    n_{1} + n_{2} + n_{3} + . . . + n_{i} = n
\end{equation}

\begin{equation}
    height_{fybrrStream} = \log_{2} n_{1} + log_{3} n_{2} + log_{4} n_{3} + . . . + log_{i+1} n_{i}
\end{equation}
Therefore we can say,
\begin{equation}
   \log_{2} n_{1} + \log_{3} n_{2} + \log_{4} n_{3} + . . . + \log_{i+1} n_{i} \leq \log_2 n
\end{equation}

Hence, height of tree in fybrrStream would always be less than or equal to height of tree in binary. Due to full utilization of stream forwarding capacity of a peer, even for a high number of live stream viewers, the height of the tree in case of fybrrStream will be quite low. 

fybrrStream follows the scoring function for each new node and assigns children to the nodes accordingly. This ensures that maximum utilization of node capacity is done. For this experiment, compared to fybrrStream, binary tree topology under-utilizes the capacity by $23.6\%$, whereas quad tree topology over-utilizes the capacity by $52.8\%$. These values can be clearly justified by the fact that node capacity is not fully utilized in case of binary, whereas nodes are over-utilized by forcing 4 children to each node in case of quad tree.
\begin{figure*}
    \centering
    \subfigure[]{\includegraphics[width=0.3\textwidth]{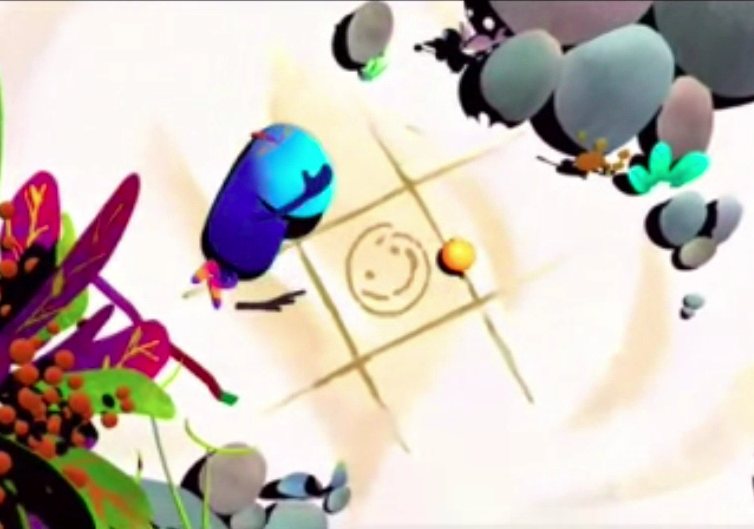}} 
    \subfigure[]{\includegraphics[width=0.3\textwidth]{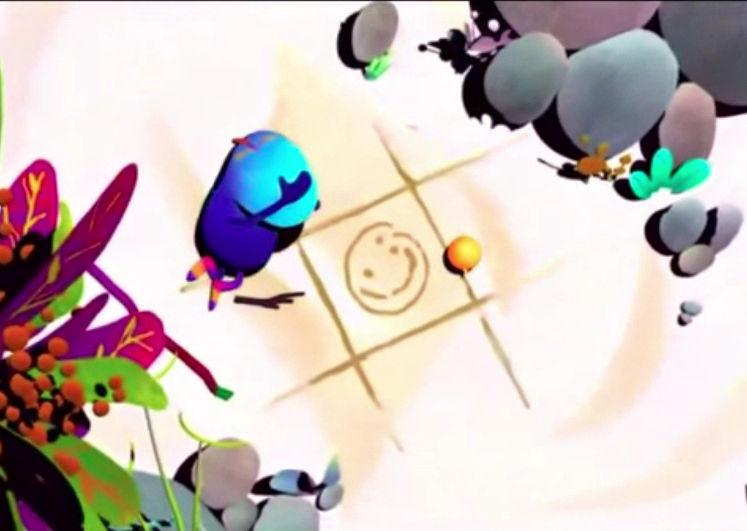}} 
    \subfigure[]{\includegraphics[width=0.3\textwidth]{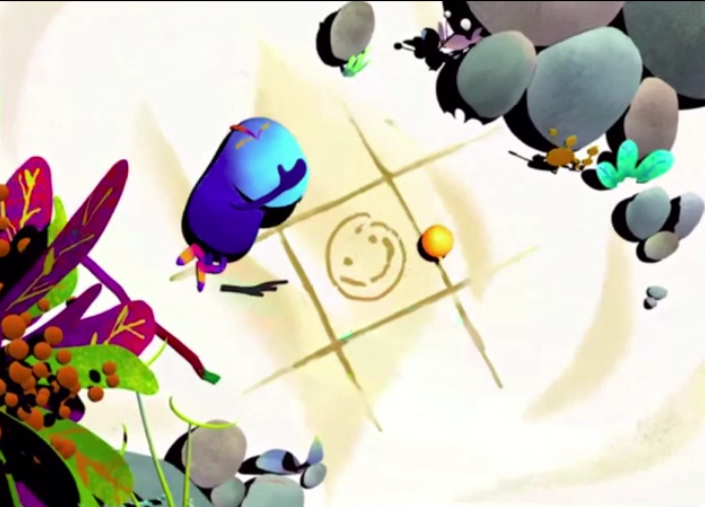}}
    \caption{Screenshot during the field testing of the schemes (a) Quad tree (b) Binary tree and (c) fybrrStream.}
    \label{fig:foobar}
\end{figure*}
%\begin{figure}[t]
	%\minipage{0.49\textwidth}
	%\includegraphics[width=\linewidth]{num-nodes.eps}
    %\caption{Number of nodes joining in the session vs %time duration from start of the session in seconds.}
    %\label{fig:num-nodes}
	%\endminipage\hfill
	~
	%\minipage{0.49\textwidth}
%   \includegraphics[width=\linewidth]{startup.eps}
 %   \caption{Startup delay (s) vs Number of nodes in the session.}
  %  \label{fig:startup-delay}
%	\endminipage\hfill

%\end{figure}
%Fig. \ref{fig:num-nodes} shows the trend for number of peers present in the session with the time duration of the session in seconds from the moment when the source node created the room for stream. A total of 56 nodes from different parts of India joined for the testing of fybrrStream.
Fig. \ref{fig:startup-delay} shows the plot for startup delay for each node joining the session. The startup delay is defined as the time taken for a node to start receiving the stream data packets from the moment when it joined the room. An average of 1.56 sec startup delay was observed for fybrrStream, which is excellent for such a system. We can also see in the graph that the startup time does not increase with increase in the number of nodes, which shows that the proposed Peer joining algorithm works well and gives stable results for any number of nodes in the system. This startup delay can be considered as a loading time for streaming services and thus we can say that a loading time of 1.5 sec is comparable or even lesser than some of the state-of-the-art streaming services which use dedicated CDN servers. Moreover, fybrrStream performs better than the proposed model in \cite{startup-comp} as well as Fast-mesh \cite{fast-mesh} in terms of the startup delay.

Fig. \ref{fig:foobar} shows the same frame of streamed video received by one of the leaf nodes in the quad, binary, and fybrrStream overlay topology. It can be seen from the images that fybrrStream provides a better quality stream than the other two schemes.
\begin{figure*}[t]
	\minipage{0.32\textwidth}
	\includegraphics[width=\linewidth]{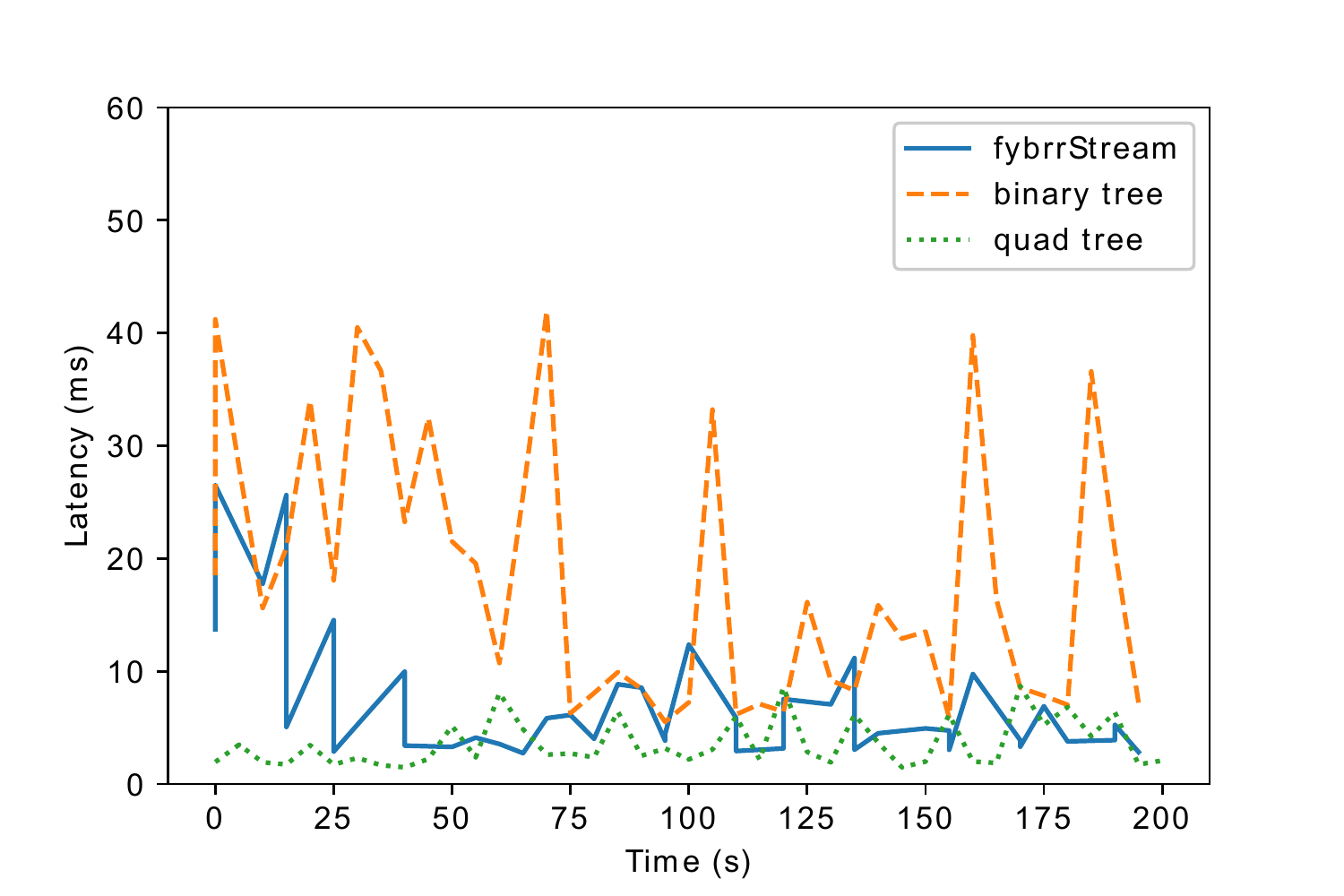}
    \caption{Average latency of leaf nodes during the stream.}
    \label{fig:latency-line}
	\endminipage\hfill
	~
	\minipage{0.32\textwidth}
   \includegraphics[width=\linewidth]{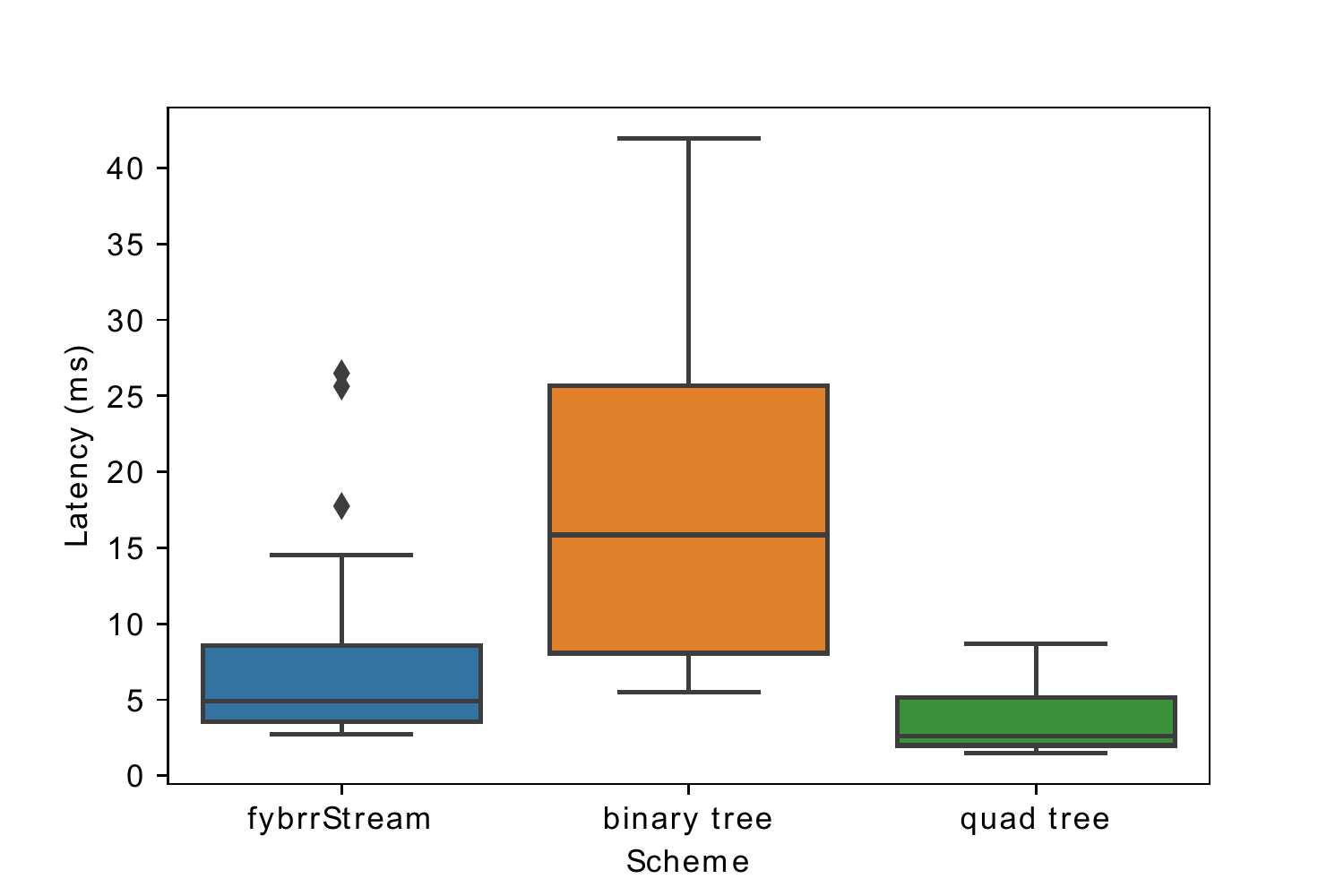}
    \caption{Average latency of leaf nodes in boxplot during the stream.}
    \label{fig:latency-box}
	\endminipage\hfill
	~
	\minipage{0.32\textwidth}
	\includegraphics[width=\linewidth]{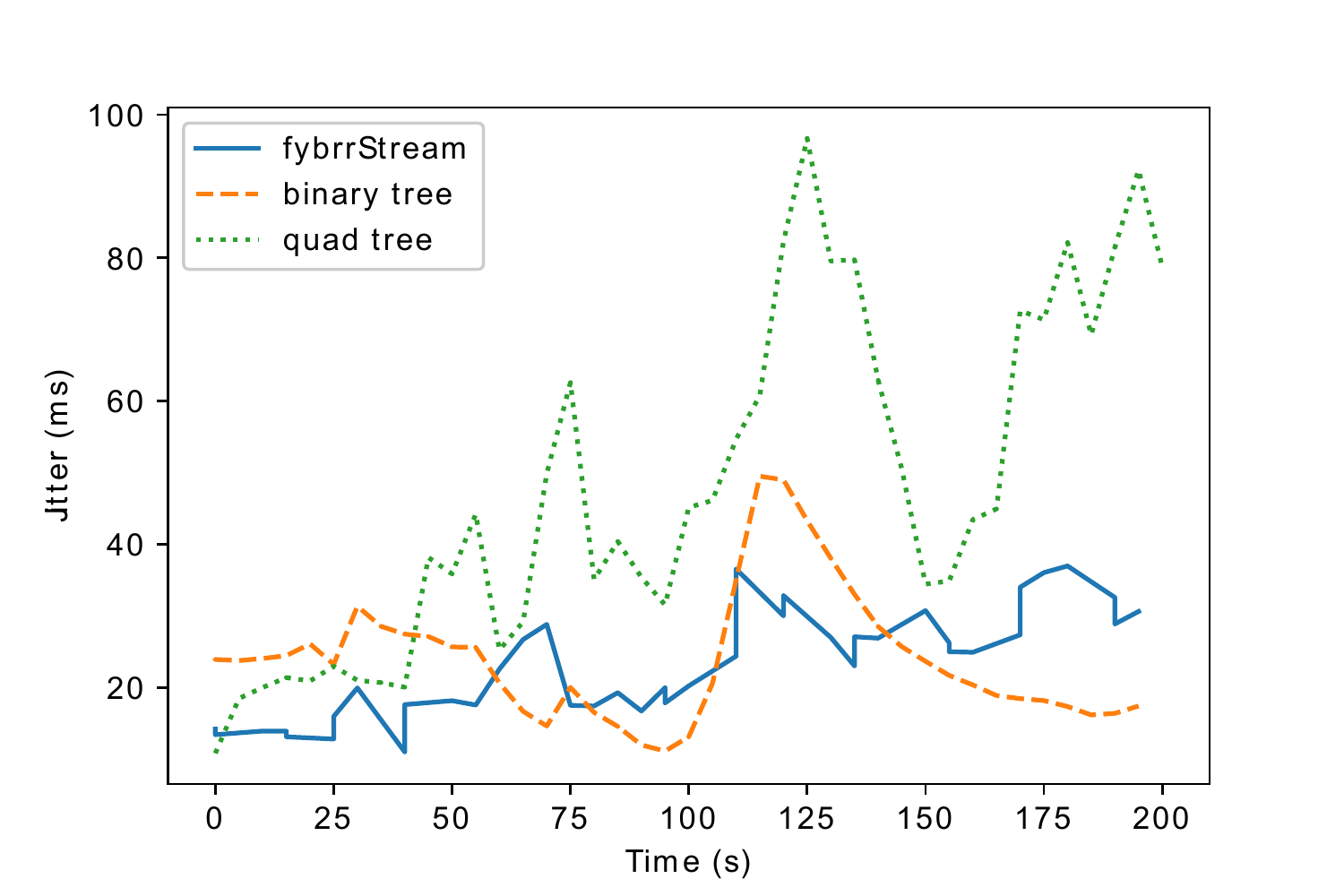}
    \caption{Average jitter of leaf nodes during the stream.}
    \label{fig:jitter-line}
	\endminipage\hfill
\end{figure*}
\begin{figure*}[t]
	\minipage{0.32\textwidth}
   \includegraphics[width=\linewidth]{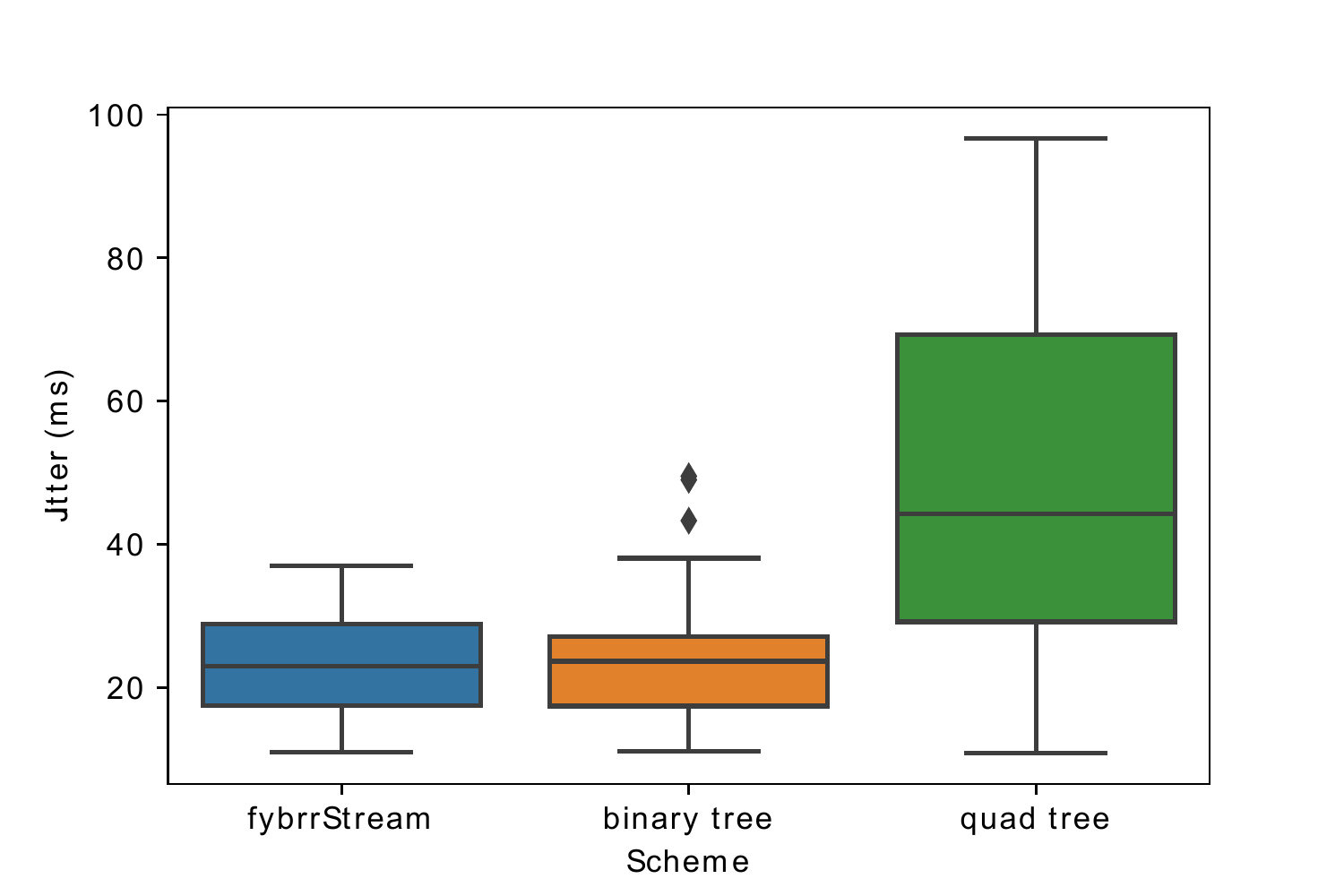}
    \caption{Average jitter of leaf nodes in boxplot during the stream.}
    \label{fig:jitter-box}
	\endminipage\hfill
	~
		\minipage{0.32\textwidth}
    \includegraphics[width=\linewidth]{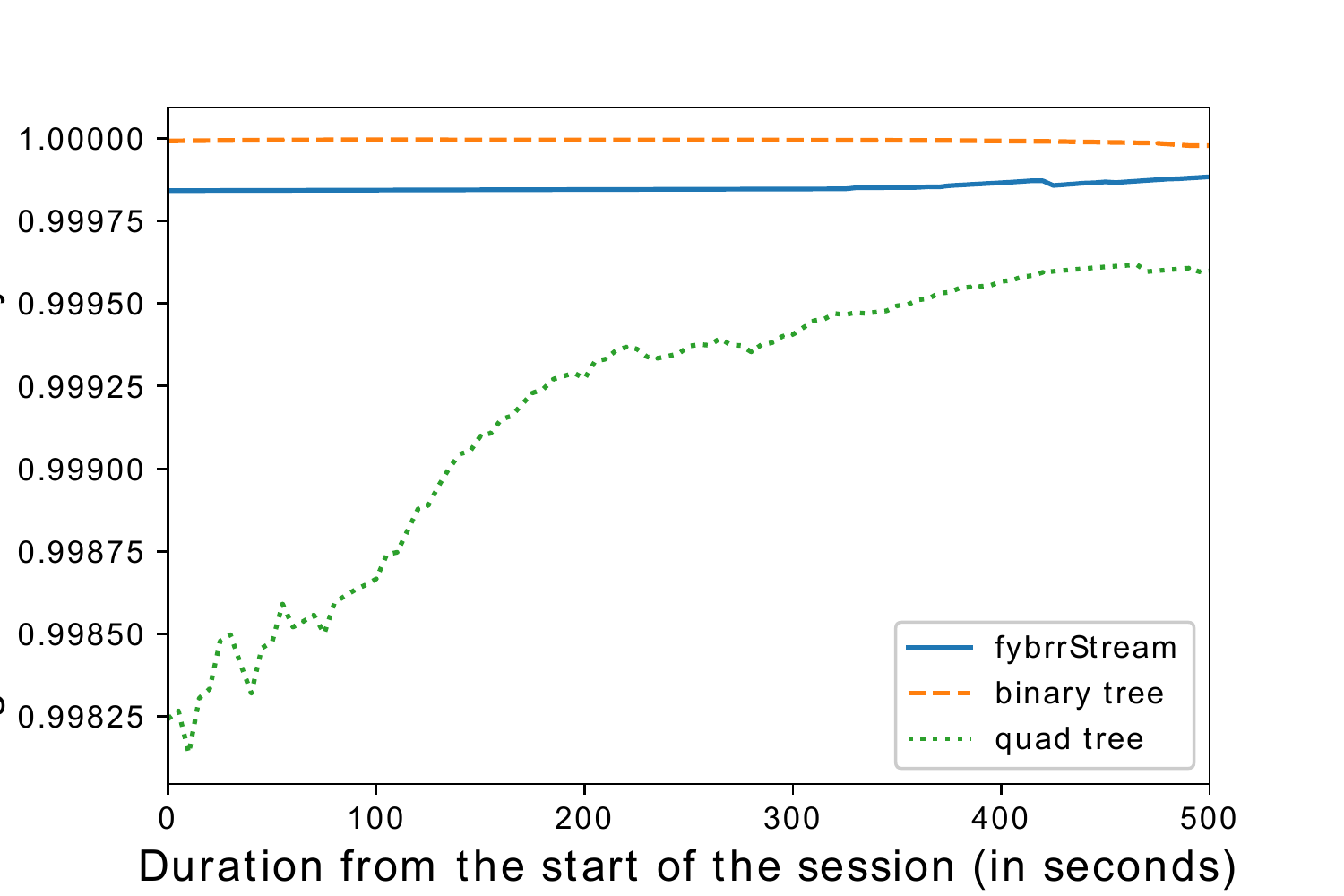}
    \caption{Average PDR achieved by the nodes vs time duration from start of the session in seconds.}
    \label{fig:qos}
	\endminipage\hfill
	~
	\minipage{0.32\textwidth}
	\includegraphics[width=\linewidth]{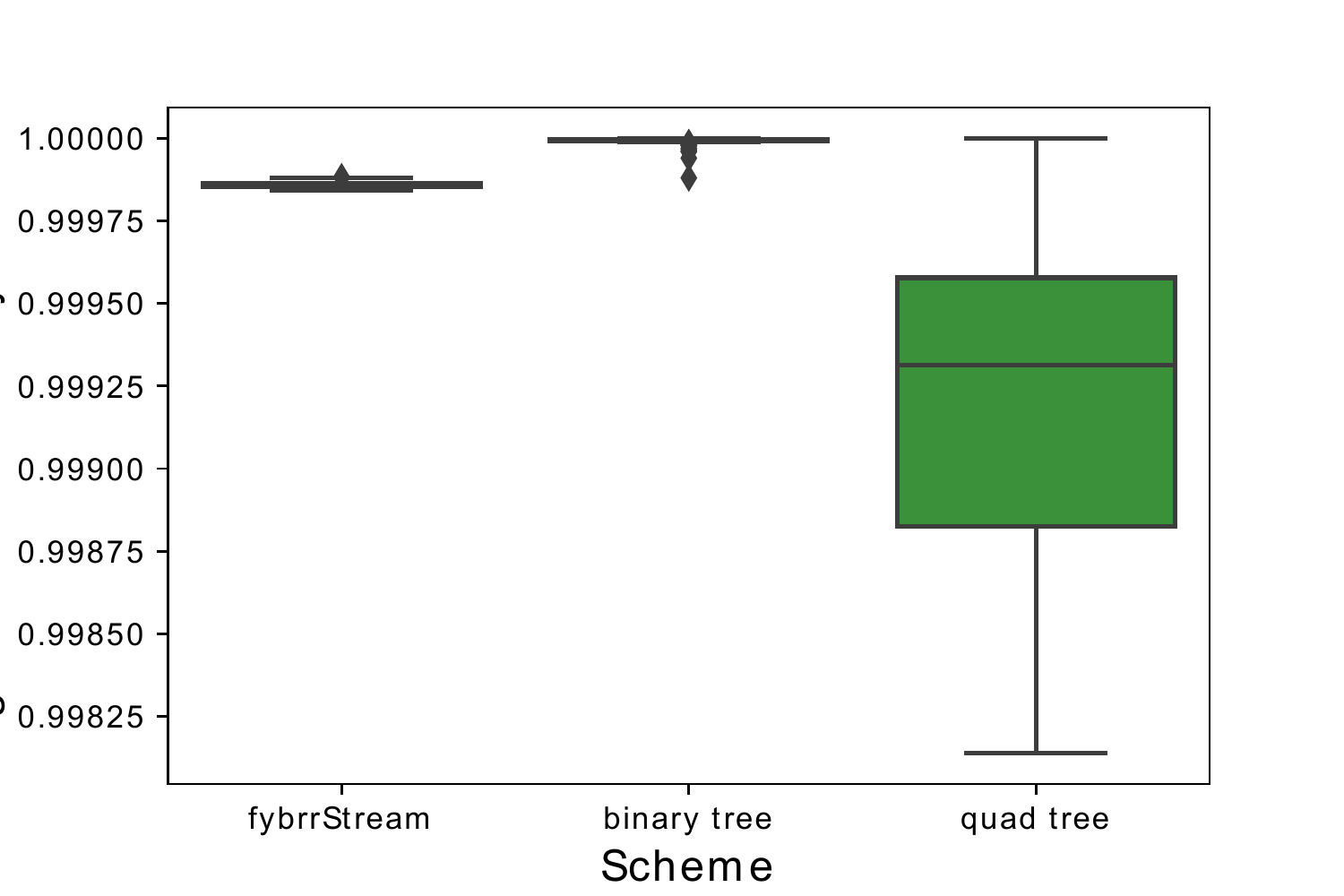}
    \caption{Boxplot for average PDR achieved by the nodes for the three schemes.}
    \label{fig:pdrbox}
	\endminipage\hfill
\end{figure*}
Fig. \ref{fig:latency-line} shows the plot for average latency at leaf nodes of the overlay network during the stream. As we see, fybrrStream gives a much lower latency than binary as it minimizes the number of hops from source node. However, at some moments, quad tree has lower latency readings than fybrrStream as it has a much lower number of hops in the network which helps to reduce the latency at leaf nodes. In Fig. \ref{fig:latency-box}, we can see that the median value of latency for all the nodes in quad tree is lowest followed by fybrrStream which is followed by binary tree. It is also evident that the median value of latency in fybrrStream is lower than minimum value obtained in the case of binary tree which suggests that fybrrStream performs much better than binary tree with respect to latency and quad performs better than fybrrStream. However, quad suffers in average PDR of the nodes (see Fig. \ref{fig:qos}) which implies if we increase number of children nodes more than what a node can support then there would be more packet loss although the latency might go down due to decrease in number of hops in the overlay~tree.

Fig. \ref{fig:jitter-line} shows the plot for average jitter at the leaf nodes with respect to the time duration of the session. As can be seen in the plot that the jitter at leaf nodes in case of quad tree is high and also has huge variation through the session whereas for binary tree case the average value is slightly higher than that of fybrrStream, which showcases that fybrrStream has a more stable network topology. Fig. \ref{fig:jitter-box} provides a more comprehensive study of the jitter. The median and the range of values for jitter in case of fybrrStream is lower than both of the other two schemes.

%\begin{figure*}[t]
%	\minipage{0.49\textwidth}
 %   \includegraphics[width=\linewidth]{all_qos.eps}
  %  \caption{Average PDR achieved by the nodes vs time duration from start of the session in seconds.}
   % \label{fig:qos}
	%\endminipage\hfill
	%~
	%\minipage{0.49\textwidth}
	%\includegraphics[width=\linewidth]{pdrbox.eps}
    %\caption{Boxplot for average PDR achieved by the %nodes for the three schemes.}
    %\label{fig:pdrbox}
	%\endminipage\hfill
%\end{figure*}

Fig.~\ref{fig:qos} shows the plot for an average PDR at any node in the session. The PDR is defined as the ratio of packets received by a node and the total number of packets transmitted for the node. The total number of packets is equal to the sum of packets received by the node and packet lost by the node. We can see in Fig. \ref{fig:qos} that the PDR for binary tree approach is highest, followed closely by the PDR of fybrrStream and then lastly by quad tree approach which is more evident in Fig.~\ref{fig:pdrbox}. It supports the fact that binary tree and fybrrStream are closely related with respect to PDR, whereas quad tree lacks behind by a significant margin, thus it is safe to conclude that fybrrStream provides sufficient PDR for better QoS. The PDR is higher in the case of a binary tree as each node only needs to forward the data packets to two other nodes. However, the height of the binary tree will be larger than fybrrStream and we would receive a higher latency, which can be seen in Fig.~\ref{fig:latency-line}. 

For fybrrStream, we assign the number of slots to a node according to its bandwidth and by this we achieve an equilibrium between latency and PDR. For the quad tree approach, the PDR is much lower, which can be explained by the fact that each node is required to forward the stream to four child nodes. Here, the quad tree approach requires a lot of uplink network bandwidth therefore the parent node is unable to match the PDR of fybrrStream and binary tree approaches.

\setlength{\extrarowheight}{1pt}
\begin{table}
 \centering
 \caption{Mean values of various performance metrics for each scheme.}
 \label{table:results}
\begin{tabular}{|p{2cm}|p{1.5cm}|p{1.5cm}|p{1.5cm}|}
\hline
Scheme & Latency (ms) & Jitter (ms) & PDR                                                                      \\ \hline \hline
fybrrStream & 11 & 15 & 0.99986 \\ \hline
Binary tree & 18 & 24 & 0.99999 \\ \hline
Quad tree & 4 & 46 & 0.99923 \\ \hline
\end{tabular}
\end{table}

Table \ref{table:results} lists the mean values of latency, jitter and PDR for all the three schemes. As can be seen in terms of latency, the binary tree has 64\% higher latency and the quad tree has 64\% lower latency than fybrrStream. Binary tree shows 60\% higher jitter than fybrrStream whereas quad tree shows 207\% higher jitter.
In terms of PDR, binary performs 0.013\% better than fybrrStream and quad performs 0.063\% worse than fybrrStream.

By this real-time field test experimentation, we tried to cover all the major factors which affect the quality of service for a user streaming through a peer-to-peer streaming service. fybrrStream worked efficiently to utilize the maximum capacity of each user node. PDR and latency plots show how the subtle choices made by the algorithm for assigning parents affect the final results whereas the comparison with the other two schemes suggests the importance of utilizing the capacity of a node efficiently.

\section{Conclusions}
\label{sec:conclusion}
The current P2P streaming platforms have shortcomings such as network complicacy in mesh, reconstruction in tree-based approaches, poor load sharing due to the heterogeneity of devices, etc., that need to be resolved for a better streaming experience. Therefore, in this paper, we proposed fybrrStream, a novel hybrid approach in which stream is pushed in tree topology and logical mesh structure is utilized for auxiliary stream forwarding. It is a WebRTC based live streaming technology that optimally distributes the load of the whole network on the participating  peers using a scoring function. The score decides the hierarchy of that peer in the overlay tree network. 
% Peers with the highest score are placed in the top levels of hierarchy in the tree as they are much more stable than the ones placed in lower levels. Parent that is less likely to fail are assigned to the new joiners. 
A quick-reconstruction algorithm is also suggested in case any parent node fails. The performance of proposed algorithms were evaluated in real-time with a field test consisting of 50+ users spread across India and results obtained showed significant improvements in the live streaming performance over binary and quad tree schemes.

%The centralized nature of proposed peer management algorithms and authorities (\emph{i.e.,} signaling server) shows the adaptability of proposed work with the next generation communication systems such as Software Defined Network (SDN) architecture \cite{6994333}. SDN controllers can take the charge of signaling and connection management for fybrrStream in an SDN environment.
\section*{Acknowledgement}
The authors thank students of 2020 batch and Infrastructure Services (ITIS) department of Indian Institute of Technology Bhilai for their contribution in the field test.

% \section*{Acknowledgment}

% The preferred spelling of the word ``acknowledgment'' in America is without 
% an ``e'' after the ``g''. Avoid the stilted expression ``one of us (R. B. 
% G.) thanks $\ldots$''. Instead, try ``R. B. G. thanks$\ldots$''. Put sponsor 
% acknowledgments in the unnumbered footnote on the first page.

% \section*{References}

% Please number citations consecutively within brackets \cite{IEEEhowto:IEEEtranpage}. The 
% sentence punctuation follows the bracket \cite{b2}. Refer simply to the reference 
% number, as in \cite{b3}---do not use ``Ref. \cite{b3}'' or ``reference \cite{b3}'' except at 
% the beginning of a sentence: ``Reference \cite{b3} was the first $\ldots$''

% Number footnotes separately in superscripts. Place the actual footnote at 
% the bottom of the column in which it was cited. Do not put footnotes in the 
% abstract or reference list. Use letters for table footnotes.

% Unless there are six authors or more give all authors' names; do not use 
% ``et al.''. Papers that have not been published, even if they have been 
% submitted for publication, should be cited as ``unpublished'' \cite{b4}. Papers 
% that have been accepted for publication should be cited as ``in press'' \cite{b5}. 
% Capitalize only the first word in a paper title, except for proper nouns and 
% element symbols.

% For papers published in translation journals, please give the English 
% citation first, followed by the original foreign-language citation \cite{b6}.

\bibliographystyle{./bibliography/IEEEtran}
\bibliography{main}

\end{document}